\documentclass[11pt]{article}
\usepackage{fullpage,latexsym,amsthm,amsfonts,amssymb,amsmath,amsthm,color,colordvi,url}

\def\01{\{0,1\}}
\newcommand{\ceil}[1]{\lceil{#1}\rceil}

\newcommand{\eps}{\varepsilon}
\newcommand{\ket}[1]{|#1\rangle}
\newcommand{\bra}[1]{\langle#1|}

\newcommand{\inpc}[2]{\langle{#1},{#2}\rangle} 
\newcommand{\norm}[1]{{\left\|{#1}\right\|}}

\newcommand{\Tr}{\mbox{\rm Tr}}

\newcommand{\E}{\mathop{\mathbb E}}

\newtheorem{claim}{Claim}

\begin{document}

\title{Optimal quantum query bounds for almost all Boolean functions}
\author{Andris Ambainis\thanks{University of Latvia, Riga. Supported by ESF project 1DP/1.1.1.2.0/09/APIA/VIAA/044.}
\and
Art\=urs Ba\v ckurs\thanks{University of Latvia, Riga. Supported by the European Commission under the project QCS (Grant No.~255961).}
\and
Juris Smotrovs\thanks{University of Latvia, Riga. Supported by ESF project 1DP/1.1.1.2.0/09/APIA/VIAA/044.}
\and
Ronald de Wolf\thanks{CWI and University of Amsterdam, rdewolf@cwi.nl. Supported by a Vidi grant from the Netherlands Organization for Scientific Research (NWO) and by the European Commission under the project QCS (Grant No.~255961).}
}
\date{}
\maketitle
\thispagestyle{empty}

\begin{abstract}
We show that almost all $n$-bit Boolean functions have bounded-error quantum query complexity at least $n/2$, 
up to lower-order terms.  This improves over an earlier $n/4$ lower bound of Ambainis~\cite{ambainis:aa}, 
and shows that van Dam's oracle interrogation~\cite{dam:oracle} is essentially optimal for almost all functions.
Our proof uses the fact that the acceptance probability of a $T$-query algorithm 
can be written as the sum of squares of degree-$T$ polynomials.
\end{abstract}

\section{Introduction}

Most known quantum algorithms have been developed in the setting of quantum query complexity,
which is the quantum generalization of the model of decision tree complexity.
Here an algorithm is charged for each ``query'' to the input bits,
while intermediate computation is free (see~\cite{buhrman&wolf:dectreesurvey} for more details about this model).
For certain specific functions one can obtain large quantum-speedups in this model.
For example, Grover's algorithm~\cite{grover:search} computes the $n$-bit OR function with $O(\sqrt{n})$ queries, 
while any classical algorithm needs $\Omega(n)$ queries.
Many more such polynomial speed-ups are known, see for example~\cite{ambainis:edj,santha:qrwsurvey,dhhm:graphproblemsj,belovs:learninggraphs}.
If one considers partial functions there are even exponential speed-ups, for example~\cite{deutsch&jozsa,simon:power,shor:factoring,bcw:sharp}.
Substantial quantum speed-ups are quite rare, and exploit very specific structure in problems that
makes those problems amenable to quantum speed-ups.

On the other hand, one can also obtain a smaller speed-up that holds for \emph{almost all} 
Boolean functions. Classically, almost all Boolean functions $f:\01^n\rightarrow\01$
have bounded-error query complexity $n$, minus lower-order terms.
This is quite intuitive: if we have only seen 99\%\ of the $n$ input bits,
then the restriction of a random function to the 1\%\ remaining variables
will still be roughly balanced between 0 and 1-inputs.
In contrast, van~Dam~\cite{dam:oracle} exhibited a beautiful quantum algorithm
that recovers the complete $n$-bit input $x$ with high probability using roughly $n/2$ quantum queries.
Briefly, his algorithm is as follows:
\begin{enumerate}
\item With $T=n/2+O(\sqrt{n\log(1/\eps)})$ and $B=\sum_{i=0}^{T}{n\choose i}$ being the number of $y\in\01^n$ with weight $|y|\leq T$, 
set up the $n$-qubit superposition 
$\frac{1}{\sqrt{B}}\sum_{y\in\01^n:|y|\leq T}\ket{y}.$
\item Apply the unitary $\ket{y}\mapsto(-1)^{x\cdot y}\ket{y}$. We can implement this using $T$ queries for $|y|\leq T$.
\item Apply a Hadamard transform to all qubits and measure.
\end{enumerate}
To see correctness of this algorithm, note that the fraction of $n$-bit strings $y$ that have weight $>T$ is $\ll\eps$.
Hence the state obtained in step~2 is very close to the state $\frac{1}{\sqrt{2^n}}\sum_{y\in\01^n}(-1)^{x\cdot y}\ket{y}$,
whose Hadamard transform is exactly~$\ket{x}$.

Since obtaining $x$ suffices to compute $f(x)$ for any $f$ of our choice,
van~Dam's algorithm implies that the $\eps$-error quantum query complexity of $f$ is
$$
Q_\eps(f)\leq n/2+O(\sqrt{n\log(1/\eps)})\mbox{ \ \ \ for all Boolean functions.}
$$
It is known that this upper bound is essentially tight for \emph{some} Boolean functions.
For example, $Q_\eps(f)=\ceil{n/2}$ for the $n$-bit Parity function~\cite{bbcmw:polynomialsj,fggs:parity}.
Our goal in this paper is to show that it is tight for \emph{almost all} Boolean
functions, i.e., that $Q_\eps(f)$ is essentially lower bounded by $n/2$ for almost all $f$
(and fixed $\eps$).  How can we prove such a lower bound?
Two general methods are known for proving quantum query lower bounds:
the polynomial method~\cite{bbcmw:polynomialsj} and the adversary method~\cite{ambainis:lowerboundsj,hls:madv}.
As we explain below, in their standard form neither method is strong enough to prove our desired $n/2$ lower bound.

First, the adversary method in its strongest incarnation~\cite[Theorem~2]{hls:madv} has the form
$$
Q_\eps(f)\geq \frac{1}{2}(1-\sqrt{\eps(1-\eps)})ADV^{\pm}(f),
$$
where the ``negative-weights adversary bound'' $ADV^{\pm}(f)$ is a quantity that is at most $n$.
Accordingly, for constant error probability $\eps$ the adversary method can only prove lower bounds of
the form $cn$ for some $c<1/2$.

Second, the polynomial method uses the fact (first proved in~\cite{fortnow&rogers:limitations,bbcmw:polynomialsj}) 
that the acceptance probability of a $T$-query algorithm can be written 
as a degree-$2T$ $n$-variate multilinear real polynomial $p(x)$ of the input.
If the algorithm computes $f$ with error probability $\leq\eps$, then $p(x)$ will approximate $f(x)$:
$p(x)\in[0,\eps]$ for every $x\in f^{-1}(0)$ and $p(x)\in[1-\eps,1]$ for every $x\in f^{-1}(1)$.
Accordingly, a lower bound of $d$ on the $\eps$-approximate polynomial degree $\deg_\eps(f)$ implies
a lower bound of $d/2$ on the $\eps$-error quantum query complexity of $f$.
This is how Ambainis~\cite{ambainis:aa} proved the current best lower bound of roughly~$n/4$ 
that holds for almost all $n$-bit Boolean functions:
he showed that almost all $f$ satisfy $\deg_\eps(f)\geq (1/2-o(1))n$.
However, O'Donnell and Servedio~\cite{odonnell&servedio:extremalj} proved a nearly matching upper bound: 
$\deg_\eps(f)\leq (1/2+o(1))n$ for almost all $f$. Hence Ambainis's lower bound approach 
via approximate degree cannot be improved to obtain our desired lower bound of $n/2$ on $Q_\eps(f)$.\footnote{In fact, 
the \emph{unbounded-error} quantum query complexity of almost all Boolean functions is only $n/4$ up to lower-order terms.  
This follows from the degree upper bound of~\cite{odonnell&servedio:extremalj} combined with \cite[Theorem~1]{bvw:smallbias} 
and the fact that $d$-bit Parity can be computed with $\ceil{d/2}$ quantum queries.}
This suggests that also the polynomial method is unable to obtain the conjectured factor 1/2 in the lower bound.

However, looking under the hood of the polynomial method, it actually gives a bit
more information about the acceptance probability: 
$p(x)$ is not an arbitrary degree-$2T$ polynomial, but the sum of squares of degree-$T$ polynomials.
Using this extra information, we prove in this paper that indeed $Q_\eps(f)\geq n/2$ 
up to lower-order terms for almost all~$f$.

\section{Proof}

Suppose we have a quantum algorithm that uses $T$ queries to its $n$-bit input $x$. 
Then by~\cite[Lemma~4.1]{bbcmw:polynomialsj}, its final state can be written as a function of the input as
$$
\sum_z \alpha_z(x)\ket{z},
$$
where $z$ ranges over the computational basis states of the algorithm's space, and the amplitudes $\alpha_z(x)$ are complex-valued multilinear $n$-variate polynomials of degree $\leq T$. We assume w.l.o.g.~that the algorithm determines its Boolean output by measuring the first qubit of the
final state. Then the acceptance probability (as a function of input $x$) is the following polynomial of degree $\leq 2T$:
$$
p(x) = \sum_{z:z_1=1} |\alpha_z(x)|^2.
$$
Let $\alpha_z\in\mathbb{C}^{2^n}$ denote the vector with entries $\alpha_z(x)$. Define the following $2^n\times 2^n$ matrix $P$:
$$
P =  \sum_{z:z_1=1} \alpha_z\alpha_z^*.
$$
The diagonal entry $P_{xx}$ of this matrix is $p(x)$. Since $P$ is positive semidefinite, we have\footnote{We use the following
matrix-analytic notation.  For $m\times m$ matrices $A$ and $B$, define inner product $\inpc{A}{B}=\Tr(A^*B)=\sum_{i,j} A_{ij}^*B_{ij}$. Note that this inner product is basis-independent: for every unitary $U$ we have $\inpc{UAU^*}{UBU^*}=\inpc{A}{B}$. Let $\norm{A}_p$ denote the (unitarily invariant) Schatten $p$-norm of $A$, which is the $p$-norm of the $m$-dimensional vector of singular values of $A$.
In particular, $\norm{A}_1$ is the sum of $A$'s singular values, and $\norm{A}_\infty$ is its largest singular value.
It is easy to see that $\norm{A}_2^2=\Tr(A^* A)=\sum_{i,j}|A_{ij}|^2$, and $\inpc{A}{B}\leq\norm{A}_1\norm{B}_\infty$.}
$$
\norm{P}_1 = \Tr(P) = \sum_{x\in\01^n} p(x).
$$
With $H$ denoting the $n$-qubit Hadamard transform, $H\alpha_z$ is proportional to the Fourier transform $\widehat{\alpha_z}$, 
which has support only on the $B = \sum_{i=0}^T{n\choose i}$ monomials of degree~$\leq T$.
Hence the matrix $HPH$ has support only on a $B\times B$ submatrix.

It will be convenient to use $+1$ and $-1$ as the range of a Boolean function, rather than 0 and~1. 
Consider Boolean function $f :\01^n\rightarrow\{\pm 1\}$. 
For $s\in\01^n$, the corresponding Fourier coefficient of $f$ is defined as $\widehat{f}(s)=\frac{1}{2^n}\sum_x(-1)^{s\cdot x}f(x)$.
Let $F$ be the $2^n\times 2^n$ diagonal matrix with diagonal entries~$f(x)$. 
Define $\widehat{F} = HFH$. Then for $s, t\in\01^n$, we have
$$
\widehat{F}_{s,t} = \bra{s}HFH\ket{t} = \frac{1}{2^n}\sum_{x,y} (-1)^{s\cdot x}(-1)^{t\cdot y}F_{xy}=\frac{1}{2^n}\sum_x (-1)^{(s\oplus t)\cdot x}f(x)=\widehat{f}(s\oplus t).
$$
Let $\widehat{F}_T$ denote $\widehat{F}$ after zeroing out all $s,t$-entries where $|s| > T$ and/or $|t| > T$. 
Note that $HPH$ doesn't have support on the entries that are zeroed out, 
hence $\inpc{HPH}{\widehat{F}} = \inpc{HPH}{\widehat{F}_T}$.

Suppose our $T$-query quantum algorithm computes $f$ with worst-case error probability at most some fixed constant~$\leq\eps$. Output~1 means the algorithm thinks $f(x)=1$, and output~0 means it thinks $f(x)=-1$. Then for every $x\in\01^n$, $2p(x)-1$ differs from $f(x)$ by at most $2\eps$. Hence:
\begin{eqnarray*}
(1 - 2\eps)2^n & \leq & \inpc{2P-I}{F}\\
 & = & 2\inpc{P}{F} - \sum_x f(x)\\
 & = & 2\inpc{HPH}{\widehat{F}} - \sum_x f(x)\\
 & = & 2\inpc{HPH}{\widehat{F}_T} - \sum_x f(x)\\
 & \leq & 2\norm{P}_1\norm{\widehat{F}_T}_\infty - \sum_x f(x)\\
 & = & 2\norm{\widehat{F}_T}_\infty\sum_x p(x) - \sum_x f(x).
\end{eqnarray*}
We can assume w.l.o.g.~that $\sum_x f(x)\geq 0$ (if this doesn't hold for $f$ then just take its negation, which has the same query complexity as $f$).
Since $\sum_x p(x)\leq 2^n$, we get 
\begin{equation}\label{eq:normFTlowerbound}
\norm{\widehat{F}_T}_\infty\geq 1/2-\eps.
\end{equation}
The technically hard part is to upper bound $\norm{\widehat{F}_T}_\infty$ for most~$f$.
So consider the case where $f:\01^n\rightarrow\{\pm 1\}$ is a \emph{uniformly random} function,
meaning that the $2^n$ values $f(x)$ are independent uniformly random signs. 
In the next subsection we show

\begin{claim}
\label{claim:maxSing}
With probability $1-o(1)$ (over the choice of $f$) we have $\norm{\widehat{F}_T}_\infty=O\left(\sqrt{\frac{n B^{1+o(1)}}{2^n}}\right)$.
\end{claim}

Combining this with the lower bound~(\ref{eq:normFTlowerbound}), we get that $B \geq 2^{n-o(n)}$.
On the other hand, a well-known upper bound on the sum of binomial coefficients is $B=\sum_{i=0}^{T}{n\choose i}\leq 2^{nH(T/n)}$,
where $H(q)=-q\log q -(1-q)\log(1-q)$ denotes the binary entropy function.
Hence, $2^{n- o(n)}\leq 2^{nH(T/n)}$ which implies $T\geq n/2-o(n)$.
This shows that $Q_{\epsilon}(f) \geq n/2-o(n)$ for almost all~$f$ (and fixed constant~$\eps$).


\subsection{Proof of Claim~\ref{claim:maxSing}}

Below, unless mentioned otherwise, probabilities and expectations will be taken over the random choice of~$f$.
We choose $T=n/2-o(n)$ sufficiently small that $B=\sum_{i=0}^{T}{n\choose i}=o(2^n)$, i.e., the $o(n)$ term in $T$ is taken to be $\omega(\sqrt{n})$.

Let $\lambda_i$ be the $i$-th eigenvalue of $\widehat{F}_T$. Since $\widehat{F}_T$ is symmetric we have
$$
\norm{\widehat{F}_T}_\infty = \max_i |\lambda_i| = \sqrt[2k]{\max_i \lambda_i^{2k}} \leq \sqrt[2k]{\sum_i \lambda_i^{2k}}=\sqrt[2k]{\Tr(\widehat{F}_T^{2k})}.
$$ 
We are going to show that 
\begin{equation}\label{eq:EFT2kupperbound}
\E\left[\Tr(\widehat{F}_T^{2k})\right] = O\left(B\left(B/2^n\right)^k\right)
\end{equation}
for every constant $k$ (with a big-O constant depending on $k$). This means that, using Markov's inequality, 
\begin{align*}
\Pr\left[ \norm{\widehat{F}_T}_\infty>C \sqrt{n B^{1+1/k}/2^n} \right] &
\leq \Pr\left[ \sqrt[2k]{\Tr(\widehat{F}_T^{2k})}>C \sqrt{n B^{1+1/k}/2^n} \right] \\
& = \Pr\left[ \Tr(\widehat{F}_T^{2k}) >C^{2k} n^k B^{k+1}/2^{nk} \right] \\
& \leq \frac{\E\left[\Tr(\widehat{F}_T^{2k})\right]}{C^{2k} n^k B^{k+1}/2^{nk}} = o(1) .
\end{align*}
Since this is true for any constant~$k$, Claim~\ref{claim:maxSing} follows.

So now our goal is to prove~(\ref{eq:EFT2kupperbound}).
Below we let each of $s_1,\ldots,s_{2k}$ range over the $B$ $n$-bit strings of weight $\leq T$, and each of $x_1,\ldots,x_{2k}$ range over $\01^n$.
For simplicity we abbreviate $\vec{s}=s_1,s_2,\ldots,s_{2k}$ and $\vec{x}=x_1,x_2,\ldots,x_{2k}$. 
Writing out the $2k$-fold matrix product, we have
\begin{align*}
\E\left[\Tr(\widehat{F}_T^{2k})\right]& =\E\left[ \sum_{\vec{s}} \widehat{f}(s_1\oplus s_2)\widehat{f}(s_2\oplus s_3)\cdots \widehat{f}(s_{2k}\oplus s_1)\right]\\
&={1 \over 2^{2nk}}\sum_{\vec{s}} \sum_{\vec{x}} \E\left[(-1)^{(s_1 \oplus s_2)\cdot x_1}f(x_1) \cdots (-1)^{(s_{2k} \oplus s_1)\cdot x_{2k}}f(x_{2k})\right]\\
&={1 \over 2^{2nk}}\sum_{\vec{s}} \sum_{\vec{x}} (-1)^{(s_1 \oplus s_2)\cdot x_1+\cdots+(s_{2k} \oplus s_1)\cdot x_{2k}}\E\left[f(x_1) \cdots f(x_{2k})\right].
\end{align*}
For a particular $y\in\01^n$, there are as many Boolean functions having $f(y)=1$ as having $f(y)=-1$, independently of what is known about values of $f$ on other inputs. Thus, if any $y$ occurs an odd number of times in $\vec{x}=(x_1, \ldots, x_{2k})$, then $\E[f(x_1) \cdots f(x_{2k})]=0$. So only those summands are left where
all multiplicities of distinct values among $x_1,\ldots,x_{2k}$ are even.
We call such $\vec{x}$ \emph{even}.
We have
\begin{eqnarray}
\E\left[\Tr(\widehat{F}_T^{2k})\right] & = & {1 \over 2^{2nk}}\sum_{\vec{s}} \sum_{\substack{\vec{x} \textnormal{ even}}} (-1)^{\sum_{i=1}^{2k}(s_i \oplus s_{i+1})\cdot x_i} \nonumber \\
& = & {1 \over 2^{2nk}}\sum_r \sum_{\substack{\textnormal{partition of }\{1,\ldots,2k\} \\ \textnormal{into even non-empty }I_1,\ldots,I_r}}\sum_{\vec{s}} \sum_{\substack{x^{(1)},\ldots,x^{(r)}\\ \textnormal{ different}}}(-1)^{\sum_{j=1}^r\left(\bigoplus_{i\in I_j}(s_i\oplus s_{i+1})\right)\cdot x^{(j)}}\label{eq:1}
\end{eqnarray}
where $s_{2k+1}=s_1$ and the second summation is over all partitions of $\{1,\ldots,2k\}$ into even-sized non-empty parts $I_1, \ldots, I_r$ with the implied condition that $x_i=x_j$ iff $i$ and $j$ belong to the same part. 
Since the number of such partitions
$(I_1, I_2, \ldots, I_r)$ depends only on $k$ (which is a constant), it suffices to prove that each term in the sum is of the order $O(B(B/2^n)^k)$. 
We will do this by proving

\begin{claim}
\label{claim:sumEst}
For any fixed $m$ and any partition $I_1,\ldots,I_r$ of $\{1,\ldots,m\}$:
\begin{equation}
\label{eq:claim}
\sum_{\vec{s}} \sum_{\substack{x^{(1)},\ldots,x^{(r)}\\ \textnormal{ different}}}(-1)^{\sum_{j=1}^r t_j(\vec{s})\cdot x^{(j)}}=O(B^{m-r+1}\cdot 2^{nr})
\end{equation}
where $t_j(\vec{s})=\bigoplus_{i\in I_j}(s_i\oplus s_{i+1})$, $s_{m+1}=s_1$, and the big-O
constant depends on $m$ and the partition.
\end{claim}

We first show that Claim~\ref{claim:sumEst} implies Claim~\ref{claim:maxSing}.
In our case, $m=2k$. Since $B=o(2^n)$, 
the upper bound $B^{2k-r+1}\cdot 2^{nr}$ increases when $r$ increases.
Since each partition of $\{1,\ldots,2k\}$ into even-sized non-empty parts $I_1, \ldots, I_r$
must contain at least 2 elements in each $I_j$, we must have $r\leq (2k)/2=k$ and
every term of the sum (\ref{eq:1}) is upper bounded by
$$
\frac{1}{2^{2nk}} O\left(B^{2k-k+1}\cdot 2^{nk}\right)=
O\left(B\left(B/2^n\right)^k\right).
$$

It remains to prove Claim~\ref{claim:sumEst}, which we do by induction on~$r$. If $r=1$ then $t_1(\vec{s})=\oplus_{i=1}^m (s_i\oplus s_{i+1})$ includes each $s_i$ exactly twice and hence sums to the all-0 string, hence
$$
\sum_{\vec{s}} \sum_{x\in\01^n}(-1)^{t_1(\vec{s})\cdot x}=\sum_{\vec{s}} \sum_{x\in\01^n}(-1)^{0\cdot x}=B^m\cdot 2^n.
$$
For the inductive step, suppose Claim~\ref{claim:sumEst} is true for $r-1$. We rewrite the left-hand side of~(\ref{eq:claim}) as
\begin{align}
\sum_{\vec{s}} & \sum_{\substack{x^{(1)},\ldots,x^{(r)}\\ \textnormal{ different}}} (-1)^{\sum_{j=1}^r t_j(\vec{s})\cdot x^{(j)}} \nonumber \\
& =\sum_{\vec{s}}\sum_{x^{(1)}}\sum_{\substack{x^{(2)},\ldots,x^{(r)}\\ \textnormal{ different}}}(-1)^{\sum_{j=1}^r t_j(\vec{s})\cdot x^{(j)}}-
\sum_{\vec{s}}\sum_{a=2}^r\sum_{\substack{x^{(2)},\ldots,x^{(r)}\\ \textnormal{ different, }x^{(1)}=x^{(a)}}}(-1)^{\sum_{j=1}^r t_j(\vec{s})\cdot x^{(j)}}.\label{eq:twosums}
\end{align}
Let us estimate both sums of~(\ref{eq:twosums}). Since $\sum_{x^{(1)}}(-1)^{t_1(\vec{s})x^{(1)}}=2^n$ if $t_1(\vec{s})=0^n$, and $=0$ otherwise, the first sum equals
\begin{equation}
\label{eq:2}
2^n\sum_{\vec{s}: t_1(\vec{s})=0}\sum_{\substack{x^{(2)},\ldots,x^{(r)}\\ \textnormal{ different}}}(-1)^{\sum_{j=2}^r t_j(\vec{s})\cdot x^{(j)}} .
\end{equation}
We now transform this sum into the form of the left-hand side of~(\ref{eq:claim}), with both $m$ and $r$ smaller by~1 compared to their current values. After that, we will apply the induction hypothesis.
 
Let $\ell$ be such that $\ell\in I_1$, $\ell-1 \notin I_1$.
Then $t_1(\vec{s})$ contains $s_\ell$ with coefficient 1 (because $t_1(\vec{s})$ includes 
$s_\ell\oplus s_{\ell+1}$ but not $s_{\ell-1}\oplus s_\ell$).
We can use the condition $t_1(\vec{s})=0$ to express $s_\ell$ in terms of $s_1, \ldots, s_{\ell-1}$
and $s_{\ell+1}, \ldots, s_m$ as follows:
\begin{equation}
\label{eq:3}
 s_\ell = s_{\ell+1} \oplus \bigoplus_{i\in I_1:i\neq \ell} (s_i \oplus s_{i+1}) .
\end{equation}
Let $b$ be such that $\ell-1\in I_b$. Then $t_b(\vec{s})$ contains $s_{\ell-1}\oplus s_\ell$
and we can substitute (\ref{eq:3}) into $t_b(\vec{s})$, obtaining
\[ 
t_b(\vec{s}) = s_{\ell-1} \oplus s_{\ell+1} \oplus \bigoplus_{i\in I_1:i\neq \ell} (s_i \oplus s_{i+1})
\oplus \bigoplus_{i\in I_b: i\neq \ell-1} (s_i \oplus s_{i+1}). 
\]
We can now remove the variable $s_\ell$ (because it was only contained in $s_{\ell-1}\oplus s_\ell$
and $s_\ell \oplus s_{\ell+1}$) and redefine $I_b$ to be $I_1\cup I_b \setminus\{\ell\}$. Then
we get that (\ref{eq:2}) is equal to
$$
2^n\sum_{\substack{s_1,\ldots,s_{\ell-1}\\s_{\ell+1},\ldots,s_m}}\sum_{\substack{x^{(2)},\ldots,x^{(r)}\\ \textnormal{ different}}}(-1)^{\sum_{j=2}^r t_j(\vec{s})\cdot x^{(j)}}
=2^n\cdot O\left(B^{m-r+1}\cdot 2^{n(r-1)}\right)=O\left(B^{m-r+1}\cdot 2^{nr}\right)
$$
with the estimate following from the induction hypothesis (with both $m$ and $r$ being smaller by 1).

As for the second sum of~(\ref{eq:twosums}), it is equal to
$$
\sum_{a=2}^r\sum_{\vec{s}}\sum_{\substack{x^{(2)},\ldots,x^{(r)}\\ \textnormal{ different}}}(-1)^{\sum_{j=2}^r t^{(a)}_j(\vec{s})\cdot x^{(j)}}=O\left(B^{m-r+2}\cdot 2^{n(r-1)}\right)
$$
where $t_j^{(a)}(\vec{s})=t_j(\vec{s})$ except for $t_a^{(a)}(\vec{s})=t_a(\vec{s})\oplus t_1(\vec{s})$ (thus merging the partition parts $I_1$ and $I_a$).
We have eliminated $x^{(1)}$ and apply the induction hypothesis (with $r$ being smaller by~1 and $m$ remaining the same).
The outer sum over $a$ introduces only a factor depending on $r\leq m$.

Since $B=o(2^n)$ we have $B^{m-r+2}\cdot 2^{n(r-1)}=o(B^{m-r+1}\cdot 2^{nr})$. Hence the bound on the first sum in~(\ref{eq:twosums}) is of a larger order and we have completed the proof of Claim~\ref{claim:sumEst}.

\bibliographystyle{alpha}

\newcommand{\etalchar}[1]{$^{#1}$}

\end{document}